# The behaviors of the wave functions of small molecules with negative kinetic energies

Huai-Yu Wang[a]

Department of Physics, Tsinghua University, Beijing, 100084 China

**Abstract:** According to relativistic quantum mechanics, particles can be of negative kinetic energies (NKE). The author asserts in his previous works that the NKE substances are dark matters. Some NKE particles, say a pair of NKE electrons, can constitute a stable system by means of the repulsive interaction between them. In the present work, two simplest three-particle systems are investigated. One consists of two NKE positrons and one NKE proton, called dark hydrogen anion. The other is composed of two NKE protons and one NKE positron, called dark hydrogen molecule cation. They are so named because the Hamiltonians of them can correspond to those of the hydrogen anion and hydrogen molecule cation. In evaluating the dark hydrogen molecule cation, the famous Born-Oppenheimer approximation does not apply, i.e., the NKE of the protons cannot be neglected. Without the NKE, the system cannot be stable. Our study reveals that in a NKE system, the particles with the same kind of electric charge combine tightly. This is to enhance the repulsive Coulomb potential so as to raise the total energy as far as possible. A great amount of NKE particles can compose a dense and dark macroscopic NKE body. Thus, it is conjectured that some remote dark celestial bodies may be NKE ones other than the well-known black holes. The discrepancies between the black holes and macroscopic NKE bodies are pointed out.

[a] wanghuaiyu@mail.tsinghua.edu.cn

**Résumé:** Selon la mécanique quantique relativiste, les particules peuvent avoir des énergies cinétiques négatives (NKE). L'auteur croit dans ses travaux antérieurs que les substances NKE sont des matières noires. Certaines particules NKE, disons une paire d'électrons NKE, peuvent constituer un système stable grâce à l'interaction répulsive entre elles. Dans le présent travail, deux systèmes à trois particules les plus simples sont étudiés. L'un est constitué de deux positrons NKE et d'un proton NKE, appelé anion hydrogène noir. L'autre est composé de deux protons NKE et d'un positron NKE, appelé cation de la molécule d'hydrogène sombre. Ils sont ainsi nommés parce que leurs hamiltoniens peuvent correspondre à ceux de l'anion hydrogène et du cation de la molécule d'hydrogène. Lors de l'évaluation du cation de la molécule d'hydrogène sombre, la célèbre approximation de Born-Oppenheimer ne s'applique pas, c'est-à-dire que le NKE des protons ne peut pas être négligé. Sans le NKE, le système ne peut pas être stable. Notre étude révèle que dans un système NKE, les particules ayant le même type de charge électrique se combinent étroitement. Il s'agit d'augmenter le potentiel de Coulomb répulsif afin d'élever l'énergie totale autant que possible. Une grande quantité de particules NKE peut composer un corps NKE macroscopique dense et sombre. Ainsi, il est supposé que certains corps célestes sombres éloignés peuvent être des corps NKE autres que les trous noirs bien connus. Les écarts entre les trous noirs et les corps NKE macroscopiques sont mis en évidence.

# I. INTRODUCTION

In Newtonian mechanics, the kinetic energy and mechanical energy are respectively defined by

$$K = \frac{p^2}{2m} \tag{1}$$

and

$$E = \frac{p^2}{2m} + V, \quad E > V, \tag{2}$$

where $V$ represents potential energy. The kinetic energy (1) is necessarily positive, called positive kinetic energy (PKE) and usually simply called kinetic energy. Consequently, in Eq. (2), the energy $E$ is impossible to be less than potential $V$.

Quantum mechanics (QM) deals with the motion of microscopic particles. Particles doing low-momentum motion obey the Schrödinger equation. The establishment of the Schrödinger equation can be regarded as replacing the momentum $p$ and energy $E$ in Newtonian mechanics by operators. The replacing operators are

$$p \to -i\hbar \nabla \tag{3}$$

and

$$E \to i\hbar \frac{\partial}{\partial t}. \tag{4}$$

By means of (3), Eqs. (1) and (2) can be expressed by operators as

$$\frac{p^2}{2m} \to -\frac{\hbar^2}{2m} \nabla^2 \tag{5}$$

and

$$\frac{p^2}{2m} + V \to -\frac{\hbar^2}{2m} \nabla^2 + V. \tag{6}$$

Both Eqs. (4) and (6) are total energies, so that they should be equal for a system. Since they are operators they have to act on a function. Letting both (4) and (6) act on a wave function $\psi$ and equating them, one obtains the Schrödinger equation.

$$i\hbar \frac{\partial}{\partial t} \psi = (-\frac{\hbar^2}{2m} \nabla^2 + V)\psi, \quad E > V. \tag{7}$$

Equation (7) is derived from (1) and (2), so that is valid for the case of $E > V$, i.e., the energy is necessarily greater than potential. Nevertheless, people found that even in the case of $E < V$, Eq. (7) was still solvable. That lead to famous tunneling effect.

However, the present author pointed out[1] that whether the Schrödinger equation was valid in the case of $E < V$ or not had neither been verified quantitatively experimentally nor derived theoretically. It is easily understood that there had been no experimental quantitative verification, because no one had detected particles inside potentials. Then, was it possible to theoretically investigate the case of $E<V$? The answer was yes.

Relativistic quantum mechanics equations (RQMEs) appeared after the Schrödinger equation. They are Klein-Gordon equation for spin-0 particles and Dirac equation for spin-1/2 particles. Theoretically, the Schrödinger equation can be regarded as a low-momentum approximation of the RQMEs. However, people did not notice that there was another low-momentum approximation,[1] which resulted in an equation other than the Schrödinger equation,

$$i\hbar \frac{\partial}{\partial t}\psi = (\frac{\hbar^2}{2m}\nabla^2 - V)\psi, \quad E<V. \tag{8}$$

Compared to the Schrödinger equation (7), there is a minus sign in the kinetic energy operator in (8), which means negative kinetic energy (NKE). Therefore, Eq. (8) stands for $E<V$, and is called NKE Schrödinger equation. As a matter of fact, Schrödinger published a series of papers when he was founding the wave equation of QM in 1926. In the fourth one of this series, he presented both (7) and (8).[2] He thought that both of them were mathematically correct. Nevertheless, he and others did not recognize the discrepancy in physics. Since then, people have been employing Eq. (7), ignoring (8).

Equation (8) is slightly different from that presented in our previous papers[1,4,6] in that the potential $V$ in (8) has a minus sign. We explain this in Appendix.

The author thinks that the two equations (7) and (8) should have the same status. That is to say, the NKE and PKE should be treated on an equal footing. Thus, all the field concerning the PKE ought to be revisited in the aspect of the NKE. It is so not only for the Schrödinger equation, but also for the RQMEs, and so a series of topics need to be researched, among which some have been done.[1,3–9] In Appendix D of [1], 13 points were listed which were topics to be dealt with in the following work. The point 1 was solved in [9], point 2 solved by [3], point 4 by [4], points 6 and 7 by [5]. Points 5 and 13 were dealt with in a submitted paper. The present paper intends to deal with point 9. The fundamental equations of QM should be in symmetric forms with respect to the PKE and NKE.[6] The virial theorem also reflected the symmetry of the PKE and NKE.[7] Up to now, all the low-momentum problems have been solved by the Schrödinger equation (7). Since Eqs. (7) and (8) apply respectively to regions of $E>V$ and $E<V$, we resolved some simple one-dimensional problems resorting to the combination of Eqs. (7) and (8).[8] The results were compared to those solved by the Schrödinger equation only, and improvements of our method were demonstrated.[8]

The point 9 in Appendix D in [1] actually addressed two topics. One topic was that that one PKE and one NKE particles could compose a stationary system, which was studied in Section 4 in [5]. The other topic was that besides the case that two NKE electrons could be paired as mentioned in Sections 3 and 4 in [1], more NKE particles with the same kind of charge could also compose a stable system by means of Coulomb repulsive interactions between them. In this work, we are going to investigate two simplest systems each being composed of three NKE particles with the same kind of charge.

Roughly speaking, the author's research idea has mainly been as follows. We first explored the properties of one NKE particle subject to different potentials by solving the NKE Schrödinger equation. [1,3–9] In this course, some related topics were studied such as solving Klein's paradox,[3] dealing with the zero-point energy of harmonic oscillator model,[9] presenting the statistical mechanics and thermodynamics of the systems composed of non-interactive NKE particles,[4] deriving the macro-mechanics from the Schrödinger equation and NKE Schrödinger equation,[5] and revisiting several one-dimensional potential problems by means of the NKE Schrödinger equation.[8] Other topics related to the NKE Schrödinger equation are that the fundamental equations in QM should be unified in that they all have the first time derivative[6] and that the Virial theorem is of symmetry with respect to PKE and NKE systems.[7] Then the two-particle systems were studied.[5] The present work investigates two simplest few-particle systems. The next work will be the formulism of many-particle NKE systems, the point 10 in Appendix D in [1]. After these works, we will proceed to scattering problems. That is to say, we will enter quantum electrodynamics (QED). The creation and destruction of particles have not been touched up to now, but they have to be involved in QED. At least two puzzles have to be clarified. One is that we have mentioned[9] that for a free particle system, there appeared a zero-point energy after second quantization, while no such an energy could be derived from the Schrödinger equation. We will show the nonexistence of this energy even after the second quantization. The other is that we will point out that in dealing with scattering problems, the NKE solutions should not be regarded as antiparticles moving in counter time direction.

One of the author's basic point of view is that the PKE substances are all what we have already known, which are composed of PKE particles, while NKE particles are dark ones. Dark particles are what we have already known. When a particle is of PKE, it can be detected by us, while when it is of NKE, it cannot, so that it is dark to us. In a submitted paper, we gave the reasons why the systems composed of NKE particles cannot be observed.

Several NKE particles can constitute a system in terms of interactions between them, just as PKE particles do. The author has researched the simplest cases, the two-body systems.[5] If the number of constituent particles increases, the wave function of the system will become complicated. When the number is few, the resultant system is called a small molecule. Here we clarify the concepts of the PKE and NKE molecules. A PKE molecule means that all the particles composing a molecule are of PKE, and electrons and nuclei have respectively negative and positive charges. There can be attractive and repulsive Coulomb interactions. A NKE molecule means that all the particles composing a molecule are of NKE and they all have the same kind of charge.

In this work, two small molecules are investigated, both consisting of three particles. In Sections II and III, a dark hydrogen anion and a dark hydrogen molecule cation are studied, respectively. The possible behaviors of the wave functions are discussed and are compare to those of corresponding PKE three-particle small molecules. Section IV is our conclusion. In Appendix, it is explained that the NKE Schrödinger equation in the form of (8) is visually better.

## II. DARK HYDROGEN ANION

First of all, we briefly review the hydrogen anion $H^-$. It is a hydrogen plus an electron. The spectral lines of the $H^-$ were first discovered in the spectra of celestial bodies.[10,11] Pekkris carefully computed its ground state energy.[12–14] The lowest value he obtained was −0.527751014 a.u. after using a complex wave function containing as many as 444 terms.[14] Chandrasekhar[15,16] proposed a wave function, called Chandrasekhar wave function, which was of simple form but relatively low ground state energy.

Let us put down the Hamiltonian of the $H^-$. The proton is located at the origin. The positon vectors of the two electrons are denoted by $r_1$ and $r_2$, respectively, and the distance between them denoted by $r_{12} = |r_1 - r_2|$. In the atomic unit, the Hamiltonian is written as

$$H_{(+)} = H_0 + H_1. \tag{9}$$

$H_0$ contains the kinetic energy of the two electrons and their attractive Coulomb potentials with the nuclear,

$$H_0 = -\frac{1}{2}\nabla_1^2 - \frac{1}{|r_1|} - \frac{1}{2}\nabla_2^2 - \frac{1}{|r_2|}. \tag{10}$$

$H_1$ is the repulsive Coulomb energy between the electrons,

$$H_1 = \frac{1}{r_{12}}. \tag{11}$$

We intend to calculate its ground state energy by wave functions which are not very complex.

The wave function is denoted by $\psi_n$, and the ground state energy is evaluated by

$$E_{(+)}(\psi_n) = \frac{\langle \psi_n | H_{(+)} | \psi_n \rangle}{\langle \psi_n | \psi_n \rangle}. \tag{12}$$

Parameters carrying radial and angular physical meanings are inserted in the wave function, and variation with respect to the parameters is taken so as to achieve the minima of the $E_{(+)}(\psi_n)$. All the trial functions are listed in TABLE I.

**TABLE I**. Energy values, in the a.u., of the trial wave functions of $H^-$ and dark $H^-$.

| $\psi_n$ | Trial wave function | Para-meters | H$^-$ | | Dark H$^-$ | |
|---|---|---|---|---|---|---|
| | | | Values of parameters | $E_{(+)}(\psi_n)$ | Values of parameters | $E_{(-)}(\psi_n)$ |
| $\psi_1$ | $e^{-r_1-r_2}$ | | | $-0.375$ | | $1.625$ |
| $\psi_2$ | $e^{-\alpha(r_1+r_2)}$ | $\alpha$ | 0.6875 | $-0.4752656$ | 1.3125 | 1.7227 |
| $\psi_3$ | $e^{-\alpha(r_1+r_2)}(1+\gamma r_{12}^2)$ | $\alpha$ | 0.865656 | $-0.5041288$ | 1.229813 | 1.743945 |
| | | $\gamma$ | 0.0758594 | | $-0.034237$ | |
| $\psi_4$ | $e^{-\alpha(r_1+r_2)}(1+\gamma r_{12})$ | $\alpha$ | 0.825726 | $-0.508780$ | 1.164523 | 1.771487 |
| | | $\gamma$ | 0.493352 | | $-0.178058$ | |
| $\psi_5$ | $e^{-\alpha(r_1+r_2)}$ $(1+\gamma_1 r_{12}+\gamma_2 r_{12}^2)$ | $\alpha$ | 0.841190 | $-0.509278$ | 1.144455 | 1.780177 |
| | | $\gamma_1$ | 0.368475 | | $-0.290785$ | |
| | | $\gamma_2$ | $-0.024660$ | | 0.025546 | |
| $\psi_6$ | $e^{-\alpha r_1-\beta r_2}+e^{-\beta r_1-\alpha r_2}$ | $\alpha$ | 1.039230 | $-0.513303$ | 1.312496 | 1.722695 |
| | | $\beta$ | 0.283221 | | 1.312497 | |
| $\psi_7$ | $(e^{-\alpha r_1-\beta r_2}+e^{-\beta r_1-\alpha r_2})$ $(1+\gamma r_{12}^2)$ | $\alpha$ | 1.0760049 | $-0.5197873$ | 1.229792 | 1.743945 |
| | | $\beta$ | 0.4882687 | | 1.229825 | |
| | | $\gamma$ | 0.0332381 | | $-0.034237$ | |
| $\psi_8$ | $(e^{-\alpha r_1-\beta r_2}+e^{-\beta r_1-\alpha r_2})$ $(1+\gamma r_{12})$ | $\alpha$ | 1.074870 | $-0.5259187$ | 1.164491 | 1.771487 |
| | | $\beta$ | 0.477447 | | 1.164493 | |
| | | $\gamma$ | 0.312549 | | $-0.178059$ | |
| $\psi_9$ | $(e^{-\alpha r_1-\beta r_2}+e^{-\beta r_1-\alpha r_2})$ $(1+\gamma_1 r_{12}+\gamma_2 r_{12}^2)$ | $\alpha$ | 1.072623 | $-0.525987$ | 1.144430 | 1.780227 |
| | | $\beta$ | 0.468525 | | 1.144424 | |
| | | $\gamma_1$ | 0.345794 | | $-0.290902$ | |
| | | $\gamma_2$ | $-0.006181$ | | 0.025610 | |

Some trial wave functions were used previously,[17] and both the three- and two-dimensional $H^-$ were studied. In these trial functions, there were at most two radial and one angular parameters, the latter being the coefficient of $r_{12}$, see TABLE I. Now, a term $r_{12}^2$ is added in $\psi_5$ and $\psi_9$. The $\psi_8$ is the Chandrasekhar wave function. The calculated results are listed in the columns with the head $H^-$, called $H^-$ columns below.

In TABLE I, the wave functions are arranged such that the $E_{(+)}(\psi_n)$ value becomes lower as the index $n$ rises. The parameter values in the $E_{(+)}(\psi_1)$ and $E_{(+)}(\psi_2)$ are analytically derived so that the two energy values are exact. Now, we analyze the behaviors of the wave functions. An intuitive consideration is that the wave functions of the two electrons should avoid each other as far as possible in order to decrease the repulsive Coulomb energies, as well as the total energy.

A trial wave function can contain two factors, each including one kind of parameters.

In a wave function, the exponential terms depend on the radial distance of the electrons and contain radial parameters. The radial parameters $\alpha$ and $\beta$ are positive, and they play roles to adjust the radial distribution of the wave function. In wave functions $\psi_n, (n=2,3,4,5)$, there is only one radial parameter $\alpha$. The greater the $\alpha$ value, the closer to the nucleus the wave function. In $\psi_n, (n=6,7,8,9)$, both $\alpha$ and $\beta$ appear, which means that radial correlation between the two electrons has been taken into account.

Another factor reflects angular correlation in the form of $1+\gamma_1 r_{12} + \gamma_2 r_{12}^2$, in which 1, $r_{12}$, and $r_{12}^2$ terms are of the physical meanings of s, p, and d partial waves, respectively. The $\gamma_1$ and $\gamma_2$ are called angular parameters. When an angular parameter is positive (negative), it reflects the effect that the wave functions of the two electrons getting departure (closer) in the angular direction.

The ground state energy of a hydrogen H is $-1/2$. When one of the radial and angular correlation is absent, the energy will be above $-1/2$, see the values of $E_{(+)}(\psi_1)$ and $E_{(+)}(\psi_2)$ in TABLE I, so that the system is not stable. In each of the functions $\psi_n, (n=3,4,5,6,7,8,9)$, there are at least two parameters. That is to say, at least one of the radial and angular correlations is considered. This makes the energy below $-1/2$, so that the system becomes stable.

In each of the functions $\psi_2$, $\psi_3$, $\psi_4$ and $\psi_5$, there is only one radial parameter and $\alpha < 1$. This discloses that the wave functions extend along the radial direction as far as possible, so as to lower the energy. Please recall that for the ground state wave function of a hydrogen atom, $\alpha = 1$.

In each of the functions $\psi_6$, $\psi_7$, $\psi_8$, and $\psi_9$, there are two radial parameters $\alpha$ and $\beta$. One of them is slightly larger than 1 and the other is remarkably less than 1. This reflects that a part of the function is closer to the nucleus and the others far away from the nucleus, showing the radial repulsive effect that the two electrons avoid each other along the radial direction as far as possible.

It is seen from TABLE I that for the $H^-$, $\gamma_1 > 0$. The p partial waves of the two electrons avoid each other along the angular direction as far as possible due to the repulsive interaction between them, in agreement with our physical intuition. The next parameter $\gamma_2 < 0$ indicates that their d partial waves slightly get closer. Please note that $\gamma_2$ is less than $\gamma_1$ by at least one order of magnitude. This is the slight adjustment against the p partial waves.

We compare $E_{(+)}(\psi_5)$ and $E_{(+)}(\psi_6)$. The former has only angular correlation and the latter only radial correlation, but the latter has a lower energy. This demonstrates that the radial correlation is more important in decreasing the energy than the angular correlation. It is easily understood that, after all, the solid angle is bounded within $4\pi$, while the radial distance can be from zero to infinity.

We turn to calculate the ground state energy of the dark hydrogen anion (dark $H^-$). A dark $H^-$ includes a static proton and two NKE positrons, all being the same kind of electric charge. The name dark $H^-$ for this system is discussable. For the time being, the author utilizes this name because comparison between the $H^-$ and dark $H^-$ can be made.

The Hamiltonian of the dark $H^-$ reads

$$H_{(-)} = -H_0 + H_1. \tag{13}$$

The first term $-H_0$, where $H_0$ is just Eq. (10), contains the NKE of the two positrons and the repulsive Coulomb potential between the proton and positrons. The second term $H_1$, Eq. (11), is the Coulomb potential between the two positrons.

The trial wave functions are the same as those of the $H^-$. The energy expectation in the function $\psi_n$ is evaluated by

$$E_{(-)}(\psi_n) = \frac{\langle \psi_n | H_{(-)} | \psi_n \rangle}{\langle \psi_n | \psi_n \rangle}. \tag{14}$$

The results are listed in the columns with the head Dark $H^-$, called Dark $H^-$

columns below.

The previous works by the author[4,5,7] pointed out that the energy spectrum of a NKE system had an upper limit but no lower limit. The higher the energy level that the system stayed, the stabler it was, a behavior just contrary to that of a PKE system. Therefore, the variational process of the evaluation of (14) is to change the values of the parameters in the wave function so as to make the calculated energy become maximum. The numerical results are shown in the Dark $H^-$ column in TABLE I. The parameter values in the $E_{(-)}(\psi_1)$ and $E_{(-)}(\psi_2)$ are analytically derived so that the two energy values are exact.

A dark hydrogen consists of a NKE positron and a proton, which was studied before,[5] and its ground state energy was $1/2$. Now, a NKE positron is added to form a dark $H^-$. If the energy of the dark $H^-$ is above $1/2$, it is a stable system.

It is seen from TABLE I that wave functions $\psi_1$ and $\psi_2$ take account of neither the radial nor angular correlation, but both the energies $E_{(-)}(\psi_1)$ and $E_{(-)}(\psi_2)$ are greater than 1/2 so that they are stable states. Even the simplest production of the wave functions of the two isolated positron, the $\psi_1$ has the energy greater than $1/2$. After the radial and angular correlations are contained, higher energies are acquired, and the dark $H^-$ becomes stabler.

Let us first examine the effect of the radial correlation. In the four wave functions $\psi_6$, $\psi_7$, $\psi_8$, and $\psi_9$, each has two radial parameters $\alpha$ and $\beta$ which can be recognized to be precisely the same. This reveals that the wave functions of the two positrons do not avoid each other. The reason is to increase their Coulomb energy as far as possible. Since these two parameters are the same, there is in fact only one radial parameter $\alpha$. Explicitly, $\psi_2 = \psi_6$, $\psi_3 = \psi_7$, $\psi_4 = \psi_8$ and $\psi_5 = \psi_9$. Indeed, in TABLE I, it is seen that $E_{(-)}(\psi_2) = E_{(-)}(\psi_6)$, $E_{(-)}(\psi_3) = E_{(-)}(\psi_7)$, $E_{(-)}(\psi_4) = E_{(-)}(\psi_8)$, and $E_{(-)}(\psi_5) = E_{(-)}(\psi_9)$. In each of these functions, the radial parameter $\alpha > 1$. Comparatively, the radial parameter $\alpha < 1$ for the wave functions of the $H^-$. This demonstrates that the wave functions of the positrons get closer to the nucleus in the dark $H^-$ than those of the electrons in the $H^-$. Near the nucleus, the higher the density of the positron clouds, the higher the repulsive Coulomb energies, and this is helpful for increasing the total energy.

We next examine the angular correlation. Contrary to the case of the $H^-$ where the coefficient $\gamma_1$ of the p partial wave is positive, the parameter $\gamma_1$ in the dark $H^-$ is negative, showing that the two positrons get closer in angular directions. This is benefit for raising the Coulomb energy, as well as the total energy. The functions $\psi_5$ and $\psi_9$ are the same, where $\gamma_1 < 0$, revealing that the p partial waves of the two positrons get close. The next parameter $\gamma_2 > 0$ shows that their d partial waves slightly depart. The value of $\gamma_2$ is less than $\gamma_1$ by one order of magnitude, reflecting a slight adjustment against the p partial wave.

In short, the wave function behavior of the dark $H^-$ is just contrary to that of the $H^-$. In the dark $H^-$, the wave functions of the two positrons have the same behavior in the radial directions and get closer to each other in the angular direction as far as possible. By comparison of the radial parameters of the $H^-$ and dark $H^-$ in TABLE I, the wave function of the latter is closer to the nucleus in the radial direction, i.e., the particles in the dark $H^-$ combine more closely. This is of cause for raising the total energy as analyzed above.

Here, we intend to make some qualitative analysis of the combinations of the constituent NKE particles in a NKE system. Before doing so, we review the factors that influence the combinations of the constituent particles in a PKE system. We always assume that the potential zero is at infinity.

A PKE system consists of PKE particles and follows energy minimum principle. Roughly speaking, the lower the total energy, the stabler the system.[4] On one hand, the attractive interactions between particles make them combine more tightly. On the other hand, the particles' PKE makes the total energy rise. The more the PKE the farther the particle is apart. If repulsive interactions between particles are added, the total energy will rise and this pushes the particles apart. The competition among the three kinds of energies make the total energy reach a minimum which is also a balance point, so that the particles compose a stationary system.

We turn to NKE systems, a case just contrary to the PKE systems. A NKE system consisting of NKE particles follows energy maximum principle.[4] Roughly speaking, the higher the total energy, the stabler the system. Thus, the more the NKE, the lower the total energy will be, and this makes the system unstable. In another word, the NKE makes the particle apart from each other. If repulsive interactions between the particles are added, the total energy of the system will rise, which is helpful for the particles to be closer. The attractive interactions between the particles, if any, will make them apart.

The competition among the three kinds of energies make the total energy reach a maximum which is also a balance point, so that the particles constitute a stationary system.

The contrary behaviors of the PKE and NKE systems are embodied in the Virial theorem already.[7]

The factors affecting the wave functions of the PKE and NKE systems are listed in TABLE II.

**TABLE II**. The factors that are in favor of particles' getting apart or getting closer in PKE and NKE systems.

|  |  | The factors in favor of particles' getting closer | The factors in favor of particles' getting apart |
|---|---|---|---|
| PKE systems |  |  | Larger particles' PKEs are in favor of particles' departure |
|  |  |  | Stronger repulsive interactions between particles are in favor of particles' getting apart |
|  |  | Stronger attractive interactions between particles are in favor of particles' getting closer |  |
| NKE systems |  |  | Larger particles' NKEs are in favor of particles' getting apart |
|  |  | Stronger repulsive interactions between particles are in favor of particles' getting closer |  |
|  |  |  | Stronger attractive interactions between particles are in favor of particles' getting apart |

In any case, kinetic energies are disadvantageous for the particles to be closer. The effects of potentials depend on if they are attractive or repulsive. A repulsive (attractive) interaction has contrary effects in a PKE and a NKE systems, as shown in TABLE II.

Two NKE electrons can form a stationary pair by means of the repulsive Coulomb interaction between them. The author proposed a method to verify such electron pairs in experiments.[1]

If there is only gravity between NKE particles, they have to disperse in space.

As concrete examples, we examine the $H^-$ and dark $H^-$, the kinetic energies and potentials of which have been clear, see Eqs. (9)-(13). The factors affecting the behaviors of the constituent particles are listed in TABLE III.

It is seen from TABLE III that for the $H^-$, there is one factor, the attraction between the nucleus and electrons, that is in favor of the electrons to be closer to the nucleus. Meanwhile, other two factors make the electrons apart as far as possible.

**TABLE III**. The factors that are in favor of the getting apart or getting closer between the particles in the $H^-$ and dark $H^-$.

|  |  | The factors in favor of particles' getting closer | The factors in favor of particles' getting apart |
|---|---|---|---|
| $H^-$ |  |  | The PKEs of the two electrons |
|  |  |  | The repulsive Coulomb interaction between the two electrons |
|  |  | The attractive Coulomb interactions between the proton and two electrons |  |
| Dark $H^-$ |  |  | The NKEs of the two positrons |
|  |  | The repulsive Coulomb interactions between the proton and two positrons |  |
|  |  | The repulsive Coulomb interaction between the two positrons |  |

As for the dark $H^-$, two factors, the repulsive Coulomb potential between the positrons and that between the nucleus and positrons, are in favor of the positrons to be closer. Only one factor, the NKE, makes the positrons apart. That is why in the dark $H^-$, the positron clouds are more closer to the nucleus compared to the electron clouds in the $H^-$. Compared to the $H^-$, the dark $H^-$ is a smaller and more compact molecule.

It is thus inferred that the more NKE particles in a system, the more closely combined the particles in order to raise the total energy as far as possible, and the shorter the averaged distances between them. It is conjectured that a great amount of NKE particles will compose a macroscopic body which is very dense, i.e., is of very high density. In principle, the density of such a system can be computed on the basis of QM as long as its Hamiltonian can be put down. The author suggests a way how to do the evaluation as the following.

Two NKE electrons can constitute a stable molecule in terms of the repulsive Coulomb potential between them.[1,6,7] This is because the competition of the NKE and positive potential is able to balance the total energy, so as to form a stationary system. The wave function of this two-particle system can be easily obtained. The mass center with the total mass of the two electrons is put at the origin, and a reduced mass moves subject to the Coulomb potential between the mass center and reduced mass, just like a hydrogen model. Now, if one more NKE electron is added, there is no doubt that the three NKE electrons will composed a stable system in terms of the repulsive potential between them. The wave function of the three-particle system can be evaluated numerically by a computer. In this way, we can add more NKE electrons one by one, and each time the wave function of the system can be calculated by computers. The diameter of each system can be judged, and the curve of the diameter vs. the electron number can be drawn. It is possible to extrapolate the curve to the case that the number is so large that the system is actually a macroscopic one. Usually, a macroscopic PKE body we can see has a clear boundary. That is to say, the thickness of the surface of a body is much less than the dimension of the body itself, and even can be regarded as zero. Based on this fact, it is inferred that when the number of the NKE electrons increases, the size of the NKE system becomes easier to determine.

It is expected that, compared with macroscopic PKE bodies, the calculated density of the macroscopic NKE bodies will be much higher.

There have been mainly three types of well-known compact celestial bodies: white dwarfs, neutron stars, and black holes, among which the black holes are dark.[18]

A black hole and a NKE body have two common remarkable features, dark and compact. Therefore, here we provide an alternative explanation of the remote dark celestials. Some of them may be the black holes generally believed, and others may be NKE bodies.

Here, we have to mention the discrepancies between the black holes and NKE bodies. 1. Although both black holes and NKE bodies are dark to us, the reasons of the darkness are different. A black hole's interior is dark because light is bounded inside the black hole and cannot leave it. While a NKE system is dark because it cannot absorb and release photons, the reason of which will be given in other work by the author. 2. A black hole is of positive temperature and pressure, while a NKE system is of negative temperature and pressure, which was analyzed in the author's previous work.[4,5] 3. The spacetime of a black hole is not Minkowski one.[18,19] A NKE body, if its mass is not very huge, is of Minkowski spacetime, just as a PKE body with a not very huge mass. If a NKE body has a sufficiently large mass, then in principle, its spacetime should be explained by the solution of Einstein's field equation. However, how to add the factor of NKE into the field equation is to be studied.

The details of the remote macroscopic NKE bodies are to be researched and the way how to research is to be explored. For instance, it may be helpful to analyze the distinctions of a body moving around a black hole and around a NKE substance. This may provide a method to distinguish if a dark celestial is a black hole or a NKE body on earth.

### III. DARK HYDROGEN MOLECULE CATION

Let us first recall the hydrogen molecule cation $H_2^+$, which is a hydrogen molecule losing an electron, i.e., two protons (called nuclei) and an electron, a three-body system. In the a.u., the Hamiltonian of the $H_2^+$ reads

$$H_{(+)} = H_e + \frac{1}{R}, \quad (15)$$

where

$$H_e = -\frac{1}{2}\nabla^2 - \frac{1}{r_a} - \frac{1}{r_b}. \tag{16}$$

The subscripts a and b label the nuclei. The $r_a = |r_a|$ and $r_b = |r_b|$ mean the distances between the electron and nuclei, and $R = |r_a - r_b|$ is the distance between the two nuclei. In Eq. (15), the first term $H_e$ includes the PKE of the electron plus its attractive Coulomb potentials with the nuclei, and the second term is the repulsive Coulomb potential between the nuclei. There were analytical solutions for the Schrödinger equation with Hamiltonian (15), so that the $H_2^+$ was believed a solvable model in QM.[20] Recently, numerical results were given for some lowest energy levels.[21]

Now we discuss how to evaluate the ground state energy. The ground state wave function of the $H_{(+)}$ is denoted by $\psi_0$. The expectation value of the $H_e$ in this function is

$$f(R) = \langle \psi_0 | H_e | \psi_0 \rangle. \tag{17}$$

The $1/R$ is a number, so that its expectation in this function is still $1/R$. The ground state energy is

$$E_{(+)}(R) = \langle \psi_0 | H_{(+)} | \psi_0 \rangle = f(R) + \frac{1}{R}. \tag{18}$$

It was analyzed[20] that $f(R)$ monotonically increased, but $1/R$ monotonically decreased, with $R$. The values of $f(R)$ at the two ends are $f(0) = -2$ and $f(\infty) = -0.5$. For the ground state energy $E_{(+)}(R)$, $E_{(+)}(0) = \infty$ and $E_{(+)}(\infty) = -0.5$, and there is a minimum at $R = 2$, $E_{(+)}(2) = -1.2$. Therefore, $R = 2$ is the distance between the two nuclei in the stationary $H_2^+$.[20]

We turn to a dark $H_2^+$. It consists of two protons and one positron, all being of NKE and the same kind of charge. Its Hamiltonian reads

$$H_{(-)} = -H_e + \frac{1}{R}, \tag{19}$$

where $H_e$ is Eq. (16). In Eq. (19), the first term $-H_e$ is the positron's NKE and the repulsive interaction between the positron and two nuclei. The second term $1/R$ in the $H_{(-)}$ is still the Coulomb energy between the nuclei. We presently use the name dark $H_2^+$ for this system, because comparisons between the $H_2^+$ and dark $H_2^+$ can be made. Whether this name is appropriate or not can be discussed.

The expectation value of the $H_{(-)}$ in the ground state wave function $\psi_0$ is

$$E_{(-)}(R) = \langle \psi_0 | H_{(-)} | \psi_0 \rangle = -f(R) + \frac{1}{R}. \tag{20}$$

As mentioned above, the function $f(R)$ increases monotonically with $R$, so that $-f(R)$ decreases monotonically. Thus, both terms of Eq. (20) decrease monotonically with $R$, and there is no maximum of $E_{(-)}(R)$ between $R=0$ and $\infty$. This leads to a conclusion: a dark $H_2^+$ cannot be stable.

However, an intuitive physical analysis prompts us that there should be a stable dark $H_2^+$. We can remove one particle out of the dark $H_2^+$. On one hand, if a nucleus is removed, the system will become a dark hydrogen which is stable. On the other hand, if the positron is removed, the system will become a pair of NKE nuclei which is also a stable dark system. In each of the two cases, as a NKE particle with the same kind of electric charge is added to form a dark $H_2^+$, the resultant ought to be a stationary system. Nevertheless, by analysis of the expectation (20), a conclusion was drawn that the system could not be stable. Why does this contradiction appear?

Let us carefully examine Eq. (15), the Hamiltonian of the $H_2^+$. When this Hamiltonian was put down, an approximation was actually made. This was famous Born-Oppenheimer approximation.[22] The reason of the approximation is that because the mass of a proton is thousands of times of that of an electron, the motion of the proton is much slower than that of the electron and cannot catch up with the latter. When one investigates the motion of the electrons, the kinetic energy of the protons can be neglected, just as if they are still. (If more precise measurement of vibration and rotation spectra of molecules is needed, the evaluation of molecules has to be more sophisticated beyond the Born-Oppenheimer approximation.[23])

The whole Hamiltonian of $H_2^+$ should include the motion of the two nuclei.

$$H_{(+)} = H_e - \frac{1}{2}\nabla_a^2 - \frac{1}{2}\nabla_b^2 + \frac{1}{R}, \tag{21}$$

where $H_e$ is still Eq. (16). Dropping the kinetic energies of the two nuclei in (21)

reduces the Hamiltonian to be (15). Hence, when we say that the Hamiltonian (15) is solvable, the premise is the ignorance of the nuclei's motion, the Born-Oppenheimer approximation. This ignorance does not bring any substantial influence to the calculation of the $H_2^+$.

For the dark $H_2^+$, however, the motion of the nuclei should not be neglected. This demonstrates that the NKE of the nuclei plays a key role to stabilize the system. All the particles in the dark $H_2^+$ are of the same kind of charge, and the potentials between them are positive. Negative energies are needed in order to make the total energy of the system reach a balance. Although the kinetic energy of the positron is negative, it is insufficient to balance the total energy, which can be recognized from the discussion of the function $f(R)$ above. Only when the nuclei's NKE is added, can the system's total energy reaches a balance. The conclusion is that the NKE of the nuclei should not be dropped. In this case, the Born-Oppenheimer approximation is not valid.

Hence, we have to put down the entire Hamiltonian of the dark $H_2^+$ as follows.

$$H_{(-)} = -H_e + H_{\text{nuclear}}, \tag{22}$$

where $H_e$ is again Eq. (16) and

$$H_{\text{nuclear}} = \frac{1}{2}\nabla_a^2 + \frac{1}{2}\nabla_b^2 + \frac{1}{R}. \tag{23}$$

Obviously, the $-H_e$ and $H_{\text{nuclear}}$ respectively describe the motion of the positron subject to the interaction of the nuclei and that of the nuclei themselves. We roughly discuss the procedure of evaluating the ground state energy of the Hamiltonian (22).

The wave function of the ground state is written as the product of those of the positron and nuclei. For a fixed $R$, the positron's wave function is denoted by $|\psi_e(R,\alpha)\rangle$, where $\alpha$ is a variational parameter. The function can take the form such as $\psi_e(\alpha) = e^{-\alpha r_a} + e^{-\alpha r_b}$. The wave function of the nuclei is denoted as $|\psi_N(\beta)\rangle$, where $\beta$ is a variational parameter. The function can take the form such as $\psi_N(\beta) = e^{-\beta R}$.

The normalization factor of the whole wave function is

$$S(\alpha,\beta) = \langle\psi_N(\beta)|\langle\psi_e(R,\alpha)|\psi_e(R,\alpha)\rangle|\psi_N(\beta)\rangle. \tag{24}$$

The ground state energy is evaluated by

$$E_{(-)} = \frac{1}{S(\alpha,\beta)}\langle\psi_N(\beta)|\langle\psi_e(R,\alpha)|H_{(-)}|\psi_e(R,\alpha)\rangle|\psi_N(\beta)\rangle. \tag{25}$$

Because the Hamiltonian (22) is composed of two parts, the ground state energy is either. One part is

$$\langle\psi_N(\beta)|\langle\psi_e(R,\alpha)|(-H_e)|\psi_e(R,\alpha)\rangle|\psi_N(\beta)\rangle$$
$$=\langle\psi_N(\beta)|g(R,\alpha)|\psi_N(\beta)\rangle \tag{26}$$

and the other is

$$\langle\psi_N(\beta)|\langle\psi_e(R,\alpha)|\psi_e(R,\alpha)\rangle H_{nuclear}|\psi_N(\beta)\rangle. \tag{27}$$

By variational method, appropriate values of the parameters $\alpha$ and $\beta$ can be obtained so as to get maximum $E_{(-)}$. In this case, we are unable to say what is the distance between the nuclei. We only can say that the averaged distance between the nuclei is computed by

$$\bar{R}=\frac{1}{S(\alpha,\beta)}\langle\psi_N(\beta)|\langle\psi_e(R,\alpha)|R|\psi_e(R,\alpha)\rangle|\psi_N(\beta)\rangle. \tag{28}$$

In the present work, we are merely content with the afore mentioned qualitative analysis of the dark $H_2^+$ that how the system can reach the stationary state physically.

By this example, we can guess an important difference between the PKE and NKE molecules.

For a PKE molecule, the kinetic energy of the nuclei can be so low that the distances between nuclei are relatively stationary. Each nucleus can do small vibration around its equilibrium position. Electrons can rotate around the connection lines between the nuclei. Consequently, a multi-nucleus molecule can have rich spectral structures containing spectral lines reflecting the states of the electrons, nuclei's vibrations, and electrons' rotations.

For a NKE molecule, the NKE of the nuclei's motion cannot be low, because it takes the role to counteract the positive energies to make the molecule reach a stable state. The nuclei move violently to have great NKE, so that the distances between the nuclei make no sense. Subsequently, in the spectra of this kind of molecules, if any, there is probably no structures reflecting the nuclei's vibrations and electrons' rotations.

## IV. CONCLUSION

In this work, two small NKE molecules are studied, and comparisons with corresponding PKE ones are made. Some qualitative conclusions of their properties are drawn.

One molecule consists of a NKE proton and two NKE positrons. It is presently named as dark hydrogen anion and its wave function is compared to that of the hydrogen anion. The clouds of the positrons are close to each other as far as possible in both the radial and angular directions, so as to raise the total energy. As a comparison, in a hydrogen anion, the two electrons' clouds avoid each other in both the radial and angular directions as far as possible to lower the total energy. It is conjectured that the more the NKE particles in a NKE system, the more tightly the particles combine so that the denser the system. The NKE substances are believed dark matters. Thus, a macroscopic NKE body is of both dark and dense features which a black hole also has. Therefore, we provide an alternative explanation of the remote dark celestials. Some dark celestials may be NKE bodies other than black holes believed nowadays. Meanwhile, the discrepancies between a black hole and a NKE body are pointed out.

The other small NKE molecule investigated is two NKE protons plus one positron, called dark hydrogen molecule cation (dark $H_2^+$) for the time being. How to calculate its wave function is discussed based on how the wave function of the hydrogen molecule cation $H_2^+$ was evaluated. In putting down the Hamiltonian of the $H_2^+$, the Born-Oppenheimer approximation is made, i.e., the motion of the atomic nuclei is neglected. However, this approximation does not apply to the dark $H_2^+$, because the NKE of the nuclei plays an important role to balance the positive Coulomb energies in the system. Without the nuclei's NKE, the dark $H_2^+$ cannot have a stationary state.

**ACKNOWLEDGMENTS**
This work is supported by the National Natural Science Foundation of China under Grant No. 12234013, and the National Key Research and Development Program of China Nos. 2018YFB0704304 and 2016YFB0700102.

**APPENDIX A: THE LOW-MOMENTUM APPROXIMATIONS OF THE DIRAC EQUATION**

The Dirac equation is

$$i\hbar \frac{\partial}{\partial t}\Psi = (c\boldsymbol{\alpha}\cdot\boldsymbol{p} + mc^2\beta + V)\Psi . \tag{A1}$$

We take transformation for the wave function.

$$\Psi = \psi_{(+)} e^{-imc^2 t/\hbar} . \tag{A2}$$

When (A2) is substituted into (A1) and low-momentum approximation is made, we obtain the Schrödinger equation,

$$i\hbar \frac{\partial}{\partial t}\psi_{(+)} = (-\frac{\hbar^2}{2m}\nabla^2 + V)\psi_{(+)} . \tag{A3}$$

Another transformation is

$$\Psi = \psi_{(-)} e^{imc^2 t/\hbar} . \tag{A4}$$

When (A4) is substituted into (A1) and low-momentum approximation is made, we obtain the negative kinetic energy Schrödinger equation,[1,4,6]

$$i\hbar \frac{\partial}{\partial t}\psi_{(-)} = (\frac{\hbar^2}{2m}\nabla^2 + V)\psi_{(-)}. \tag{A5}$$

For a given potential $V$, if the Schrödinger equation (A3) has solutions, then, the condition that (A5) also has solutions is that the potential in (A5) should be just minus of that in (A3).

We think the Schrödinger equation and the NKE Schrödinger equation should be of symmetric forms for a given potential $V$, but Eqs. (A3) and (A5) are not.

As a matter of fact, we have pointed out[1] that the $\boldsymbol{\alpha}$ in (A1) can have a minus sign.

$$i\hbar \frac{\partial}{\partial t}\Psi = (-c\boldsymbol{\alpha}\cdot\boldsymbol{p} + mc^2\beta + V)\Psi. \tag{A6}$$

Equation (A6) is still the Dirac equation. Now, we substitute (A4) into (A6) and then take low-momentum approximation. The resultant is the following NKE Schrödinger equation.

$$i\hbar \frac{\partial}{\partial t}\psi_{(-)} = (\frac{\hbar^2}{2m}\nabla^2 - V)\psi_{(-)}. \tag{A7}$$

Equations (A3) and (A7) are derived from one Dirac equation. Therefore, they have the symmetric forms as should be. For instance, for a hydrogen atom, Dirac equation with Coulomb potential has PKE and NKE solutions.[1] The PKE solutions are from the attractive potential and the NKE solutions are from the repulsive potential. Under low-momentum approximations, the PKE and NKE solutions should be respectively resolved from (A3) and (A7). That the $\boldsymbol{\alpha}$ takes a minus sign, as in (A6), actually reflects the exchange of the PKE and NKE solutions.

At last, we have some more words comparing Eqs. (A5) and (A7). They are essentially the same in that both have negative kinetic energies, because they both are derived from the Dirac equation by means of transformation (A4). Equation (A7) is visually more comfortable since the whole right hand side is just the minus sign of the right hand side of the Schrödinger equation (A3). The form (A7) stresses that the potential $V$ in this equation is exactly the same as that in (A3) for both equations are obtained from the Dirac equation (A1) with this $V$. The form (A5) means that from the Dirac equation we can obtain an NKE Schrödinger equation with a potential which is not specified. Therefore, the difference between (A5) and (A7) that they have the counter sign in potential does not bring substantial problems. The key point is the NKE in the equations. The potential can vary depending on concrete system under investigation. All the discussion and conclusions in the author's previous works[1,3–9] are not affected when Eq. (A5) is replaced by (A7), because what the author have studied are the physical properties of NKE systems.

X **7**, 031035 (2017). https://10.1103/PhysRevX.7.031035